\documentclass[10pt,letterpaper,twocolumn,prl,amsmath,amssymb,nobalancelastpage,aps,showpacs,preprintnumbers]{revtex4-1}
\usepackage[utf8]{inputenc}
\usepackage{bm}
\usepackage{amsmath}
\usepackage{amssymb}
\usepackage{graphicx}

\makeatletter

\pdfpageheight\paperheight
\pdfpagewidth\paperwidth

\usepackage{dcolumn}%
\usepackage{bm}%
\usepackage{slashed}
\usepackage{amsthm}
\usepackage{youngtab}
\usepackage{color}

\makeatother

\begin{document}

\title{Strong zero modes from geometric chirality in quasi-one-dimensional Mott insulators}

\author{Raul A. Santos$^{1}$ and Benjamin B\'eri$^{1,2}$}

\affiliation{$^{1}$T.C.M. Group, Cavendish Laboratory, University of Cambridge, J.J. Thomson Avenue, Cambridge, CB3 0HE, United Kingdom}

\affiliation{$^{2}$DAMTP, University of Cambridge, Wilberforce Road, Cambridge, CB3 0WA, United Kingdom}

\begin{abstract}
Strong zero modes provide a paradigm for quantum many-body systems to encode local degrees of freedom that remain coherent far from the ground state. 
Example systems include $\mathbb{Z}_n$ chiral quantum clock models with strong zero modes related to $\mathbb{Z}_n$ parafermions.
Here we show how these models and their zero modes arise from geometric chirality in fermionic  Mott insulators, focusing on $n=3$ where the Mott insulators are three-leg ladders. 
We link such ladders to $\mathbb{Z}_3$ chiral clock models by combining bosonization with general symmetry considerations.
We also introduce a concrete lattice model which we show to map to the $\mathbb{Z}_3$ chiral clock model, perturbed by the Uimin-Lai-Sutherland Hamiltonian arising via superexchange.
We demonstrate the presence of strong zero modes in this perturbed model by showing that correlators of clock operators at the edge remain close to their initial value for times exponentially long in the system size, even at infinite temperature.
\end{abstract}

\maketitle

Quantum many-body systems supporting local degrees of freedom that remain coherent for long times even away from the ground state may open up nonzero temperature, or even non-equilibrium, regimes for quantum information processing~\cite{Fendley2012,Fendley2014,Fendley2016,Alicea2016,Kemp2017,Else2017,Kemp19,Yates20,WoottonPRL2011,StarkPRL2011,2013Bauer_Nayak,Huse2013LPQO,Parameswaran2017,Alet2017,Parameswaran2018,Serbyn2019,Wahl20}. 
A key requirement for such coherence is the presence of degeneracies across the energy spectrum.
``Strong zero modes"~\cite{Fendley2016} 
provide a compelling mechanism for this (another route is by many-body
localization~\cite{WoottonPRL2011,StarkPRL2011,2013Bauer_Nayak,Huse2013LPQO,Parameswaran2017,Alet2017,Parameswaran2018,Serbyn2019,Wahl20}):  
these are objects that commute with the Hamiltonian (up to corrections exponentially decaying in system size), but they do not commute with a discrete symmetry, %
hence ensuring spectral degeneracies. 
When located at the system edge, they furthermore furnish the desired long-time-coherent local degrees of freedom~\cite{Fendley2012,Fendley2014,Fendley2016,Alicea2016,Kemp2017,Else2017,Kemp19,Yates20}.  %

One of the most intriguing paradigms where zero modes appear are quantum clock models with $\mathbb{Z}_{n\geq 3}$ symmetry~\cite{Fendley2012,Fendley2014,Fendley2016,Alicea2016}.
The zero modes include $\mathbb{Z}_n$-parafermions,
signifying a (nonlocal) relation to electronic systems proposed for beyond--Majorana schemes of topological quantum computation~\cite{Lindner2012,Cheng2012,Clarke2013,Mong2014,Tsvelik2014,Klinovaja2014PRL,Zhang2014,Orth2015,Alexandradinata2016,Alavirad2017,Calzona2018,Mazza2018,Fendley2012,Fendley2014,Fendley2016,Hutter2016,Alicea2015,Alicea2016,Chew2018,Santos2018}.
To support strong zero modes, the clock models require a chiral (i.e., reflection-symmetry breaking) deformation of their couplings~\cite{Fendley2012,Fendley2014,Fendley2016,Alicea2016,Jermyn2014,Zhuang2015,Moran2017}.  
While phase transitions in the chiral quantum clock model universality class have seen realizations~\cite{Fendley2004,Lecheminant2012,tsvelik2012parafermion,Samadjar2018,keesling2019quantum}, the chiral quantum clock models themselves, and their strong zero modes, are yet to find their origin in an underlying microscopic system.

Here we describe a paradigm for how chiral quantum clock models and their strong zero modes can arise in  Mott insulators. 
While somewhat abstract in terms of clock models, 
chirality is a simple geometrical feature
for 
particles hopping on a lattice~\cite{buda1992quantifying}, with examples such as chiral nanotubes, molecules, or crystals~\cite{dresselhaus1998physical,fasman2013circular}. 
Our approach is centered on the combination of such geometric chirality with strong interactions. 

\begin{figure}
 \includegraphics[width=0.9\linewidth]{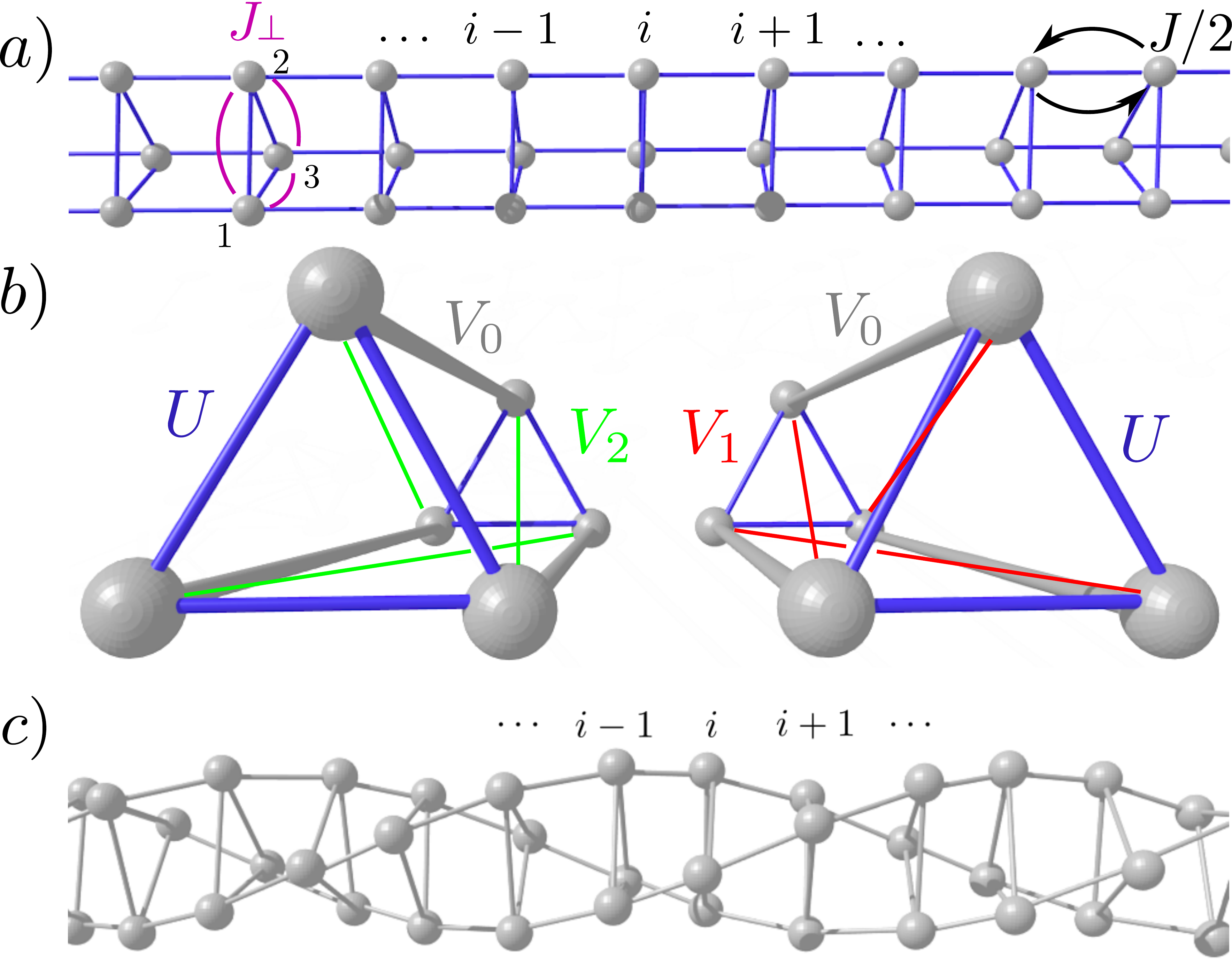}
\caption{Three-leg geometry for spinless fermions. a.- The 
hopping amplitudes $J$ and $J_{\perp}$.
b.- The interaction $U$ between fermions at the same $i$ and the interactions between neighboring sites $V_0,V_1$ and $V_2$. 
Geometric chirality (e.g., as in panel c) naturally leads to chiral interactions ($V_1\neq V_2$). 
}
\label{fig:Single_particle_system_layout}
\end{figure}

We focus on the simplest case of $\mathbb{Z}_3$ symmetry and study spinless fermion systems such as the three-leg ladder in Fig.~\ref{fig:Single_particle_system_layout}. 
We take two complementary approaches: (i) bosonization that captures generic features beyond a single microscopic model but which is only phenomenological in interactions, and (ii) strong-interaction perturbation theory for the system in Fig.~\ref{fig:Single_particle_system_layout}. 
Using bosonization, we show how chiral-quantum-clock-model physics, including zero modes, can arise in the presence of certain symmetries (e.g., $\mathbb{Z}_3$ and time-reversal) and sufficiently strong and chiral interactions. 
In our lattice system, we provide an explicit mapping to the chiral three-state clock model perturbed by terms arising via superexchange,
and assess the presence of strong zero modes by computing dynamical correlators of clock operators at the edge using exact diagonalization.

{\it Models and symmetries}.-\label{sec:model}
We set our terminology using the system in Fig.~\ref{fig:Single_particle_system_layout}. 
Anticipating the clock-model mapping, we refer to the positions $i$ along the legs as sites. 
The legs are of length $\ell$ and the sites are $a_0$ apart. 
The Hamiltonian is $H=H_0+H_\text{int}$ with
\begin{eqnarray}\label{Ham_lattice}\nonumber
H_0&=&-\frac{J}{2}\sum_{i,a}c_{i,a}^{\dagger}c_{i+1,a}+J_\perp\sum_{i,a}c_{i,a}^{\dagger}c_{i,a+1}+\text{H.c.},\\
H_\text{int}&=&\sum_{i,a,b}\left(V_b\, n_{i,a}n_{i+1,a+b}+\frac{U}{2}\,n_{i,a}n_{i,b}\right).
\end{eqnarray}
Here $c^\dagger_{i,a+3}=c^\dagger_{i,a}$ creates a fermion at site $i$ and leg $a\in\{0,1,2\}$;
$n_{i,a}=c_{i,a}^{\dagger}c_{i,a}$ is the corresponding number operator.
We use real intraleg and interleg tunneling amplitudes, $J$ and $J_{\perp}$, respectively.

Geometric chirality is present if $V_1\neq V_2$. We characterize this by introducing $V=\frac{1}{3}\sum_aV_{a}$ and 
\begin{equation}\label{parameters_chiral}
V'\sin\Phi =\frac{V_{1}-V_{2}}{\sqrt{3}},\quad V'\cos\Phi =\frac{2V_{0}-V_{1}-V_{2}}{3}.
\end{equation}
The system is invariant under the $\mathbb{Z}_3$ transformation $\mathcal{S}$ cyclically permuting the legs. 
Further symmetries include time-reversal $\mathcal{T}$ and, for the bulk physics (i.e., far from the boundaries), lattice translations along the legs.
The generic systems we shall discuss are three-leg ladders beyond Eq.~\eqref{Ham_lattice}, but which still respect these symmetries~\cite{symmfn}.
We work at $1/3$ filling, $\sum_{i,a} n_{i,a} a_0/\ell=1$. 

\begin{figure}
 \includegraphics[width=\linewidth]{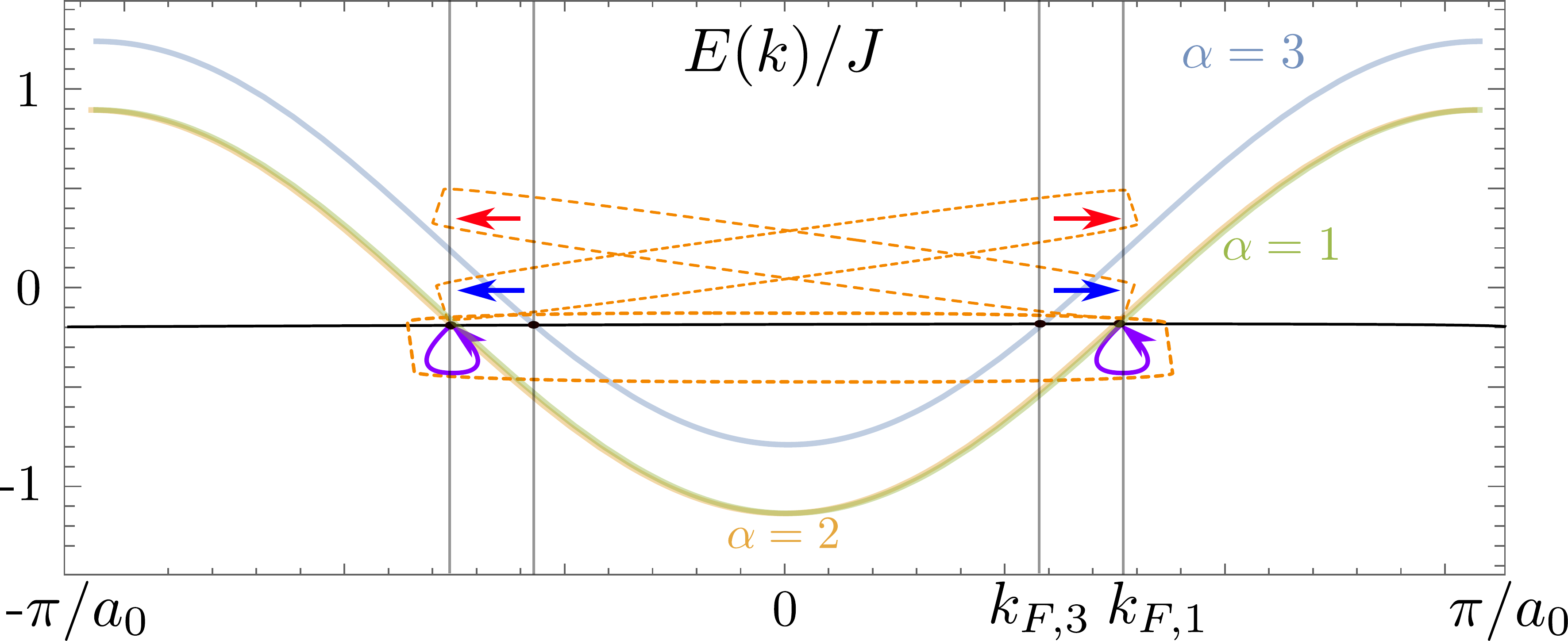}
\caption{The single-particle spectrum of $H_0$ in Eq.~\eqref{Ham_lattice} for $J_\perp/J=0.12$. 
Bands $\alpha=1,2$ remain degenerate for nonzero $J_\perp$~\cite{kFfn}. 
At $1/3$ filling, momentum-conserving interband four-fermion processes combine pairs of two-fermion processes (arrows) between $k_{F,\alpha}$. The arrow colors indicate the bands involved (red: 3 and 1; blue: 3 and 2; purple: 1 and 2); the vertical offset is only for visualization. 
$\mathbb{Z}_3$ symmetry fixes the combinations to be those in the dashed boxes.
Reversing all arrows gives the Hermitian conjugate processes.
}
\label{fig:Single_particle_system_disp}
\end{figure}

{\it Low-energy processes}.-
We next prepare for bosonization by describing the low-energy bulk processes. 
By low-energy, we mean processes near the Fermi points. 
Our bosonization thus starts with moderate interactions; however, its phenomenological scope is broader and includes strong interactions~\cite{GiamarchiBook2003,GogolinBook2004}.
We first diagonalize the $\mathbb{Z}_3$ transformation using $f_{i,\alpha}=\sum_{a}v_{\alpha a}c_{i,a}$, where $v$ is a $3\times3$ unitary matrix; this diagonalizes the single-particle Hamiltonian into three bands labeled by  $\alpha\in\{1,2,3\}$ (Fig.~\ref{fig:Single_particle_system_disp} for our concrete model~\cite{kFfn}). 
$\mathbb{Z}_3$ and time-reversal symmetries now act as
$\mathcal{S}f_{j,\alpha}\mathcal{S}^{-1}= \omega^\alpha f_{j,\alpha}$ (where 
\mbox{$\omega=e^{2\pi i/3}$),} and \mbox{$\mathcal{T}f_{j,1}\mathcal{T}^{-1}=f_{j,2}$,} 
$\mathcal{T}f_{j,3}\mathcal{T}^{-1}=f_{j,3}$. 
At $1/3$ filling, and for moderate band splitting [$J_\perp\lesssim J/3$ for Eq.~\eqref{Ham_lattice}], there are six Fermi points $k_{F,\alpha}$. 
Working near $k_{F,\alpha}$, we can split $f_{j,\alpha}$ into left ($L$) and right ($R$) movers, 
\mbox{$\frac{f_{j,\alpha}}{\sqrt {a_0}}=R_{\alpha}(x)e^{ik_{F,\alpha} x}+L_{\alpha}(x)e^{-ik_{F,\alpha} x},$}
with $x=j a_0\in[0,\ell]$.
The %
symmetries (including crystal-momentum conservation)
allow three classes of four-fermion processes:  
forward scattering, band-$3$ pairing, and 
``not-$3$'' scattering.
Forward scattering contributes to the quadratic part of the theory~\cite{GiamarchiBook2003,GogolinBook2004}. 
The band-$3$ pairings $O_{\text{p}1}=L_{1}^{\dagger}R_{2}^{\dagger}L_{3}R_{3}$ and $O_{\text{p}2}=L_{2}^{\dagger}R_{1}^{\dagger}L_{3}R_{3}$ 
describe the transfer between band-$3$ fermion pairs and fermions in bands $1$, $2$. 
Not-$3$ scattering $O_{\bar{3}}=L_{1}^{\dagger}R_{2}^{\dagger}L_{2}R_{1}$  does not involve band-$3$ fermions. 
These, and their Hermitian conjugates, are the lowest-order symmetry-allowed processes. 
They transfer R-movers (and L-movers) across different bands, but conserve their total number separately. 
To capture the phenomenology, we include the umklapp $O_\text{u}=L_1^\dagger L_2^\dagger L_3^\dagger R_1 R_2 R_3$, which is the lowest-order symmetry-allowed process scattering R- and L-movers into each other without inter-band transfer.

{\it Bosonization}.- The complementary character of the $O_{a\neq \text{u}}$ and $O_{\text{u}}$ processes translates to the separation into charge and neutral degrees of freedom. 
This becomes transparent in bosonization, an approach describing the corresponding density fluctuations~\cite{GiamarchiBook2003,GogolinBook2004}.
We consider densities for charge $\tilde{N_c}=\sum_j N_j$ and neutral $\tilde{N_1}=N_1-N_2$, $\tilde{N_2}=N_1+N_2-2N_3$ combinations, where $N_\alpha$ is the particle number in band $\alpha$. Note that $\mathcal{S}=\exp\left(\frac{2\pi i}{3}\tilde{N_1}\right)$, as implied by the transformation of $f_\alpha$.
We use conjugate pairs of fields $\theta_\mu$ and $\varphi_\mu$ ($\mu=c,1,2$) whose
only nonzero commutators are $[\theta_\mu(x),\varphi_\nu(y)]=i\delta_{\mu\nu}\Theta(x-y)$ with $\Theta(x)$ the Heaviside step function. 
The densities (relative to the Fermi sea) are $\rho_\mu = \sqrt{\beta_{\mu}/\pi}\,\partial_x \theta_\mu$ with $\beta_{c,1,2}=3,2,6$. 
In terms of these fields, the band-$\alpha$ electron operators $\psi_{+\alpha}=R_\alpha$ and $\psi_{-\alpha}=L_\alpha$ are
$\psi_{\eta\alpha}\propto \frac{\kappa_{\alpha}}{\sqrt{a_{0}}}\exp\left\{ i\sqrt{\pi}\left[(\varphi_{c}+\eta\theta_{c})/\sqrt{3}+\bm{d}_{\alpha}\cdot(\tilde{\bm{\varphi}}+\eta\tilde{\bm{\theta}})\right]\right\}$
where $\kappa_{\alpha}$ are Klein factors, 
$\tilde{\bm{\varphi}}=(\varphi_{1},\varphi_{2})$, 
$\tilde{\bm{\theta}}=(\theta_{1},\theta_{2})$, 
$\bm{d}_{1}=(1/\sqrt{2},1/\sqrt{6})$,
$\bm{d}_{2}=(-1/\sqrt{2},1/\sqrt{6})$, $-\bm{d}_{3}=\bm{d}_{1}+\bm{d}_{2}$~\cite{Lecheminant2012,Teo2014,Fabrizio}. 
At the boundaries ($x=0,\ell$), zero charge current implies $\partial_x\varphi_c|_{x=0,\ell}=0$; in the neutral sector we require only that boundary conditions respect $\mathcal{S}$ and $\mathcal{T}$.

The low-energy bulk Hamiltonian is $H_\text{b}=H_{\rm fw} + \int dx \mathcal{V}$. 
Here $H_{\rm fw}$ encodes single-particle and forward-scattering terms. 
Although $H_{\rm fw}$ influences the phase diagram, for the essential features of the gapped regime we are interested in we can focus entirely on 
\begin{multline}\label{potential}
\mathcal{V}(\theta_c,\theta_1,\varphi_2)=g_{\bar{3}}\cos\sqrt{8\pi}\theta_{1} +g_2\cos\sqrt{2\pi}\theta_{1}\cos\sqrt{6\pi}\varphi_{2}\\
+g_1\sin\sqrt{2\pi}\theta_{1}\sin\sqrt{6\pi}\varphi_{2} + g_\text{u}\cos\sqrt{12\pi}\theta_c
\end{multline}
encoding the $O_a$ processes. 
(We absorbed in $g_{1,2}$ the product $i\kappa_1\kappa_2$.)
The following hold for the neutral-sector couplings $g_{1,2,{\bar{3}}}$, regardless of microscopic details: 
(i) interactions involving only the total
density [such as the $U$ term in Eq.~\eqref{Ham_lattice}] conserve each band particle number and hence do not contribute to $g_{1,2}$;
(ii) chirality enters via $g_1$ because spatial reflections take $O_{\text{p1}}\leftrightarrow O_{\text{p2}}$ and preserve $O_{\bar{3}}$, so $g_{2,\bar{3}}$ are invariant but $g_1$ changes sign. 
The physics we are interested in is where $\mathcal{V}$ has deep minima confining $\theta_{c,1}$ and $\varphi_2$, 
a gapped regime that is expected to arise for sufficiently strong repulsive total-density interactions, as we shall confirm for our concrete model Eq.~\eqref{Ham_lattice}. 
Such strong microscopic interactions may place the problem beyond the scope of 
the weak-coupling renormalization group. 
One can, however, seek the chiral clock model phenomenology via a semiclassical analysis~\cite{GiamarchiBook2003,GogolinBook2004} of the field configurations that minimize $\mathcal{V}$.
We start with the charge sector:
the field $\theta_c$ is a constant locked to one of the minima of the umklapp cosine. 
The fluctuations $\rho_c\propto \partial_x \theta_c$ are thus absent; $1/3$-filling now means  constant density of one particle per site. 

\begin{figure}
 \includegraphics[width=0.8\columnwidth]{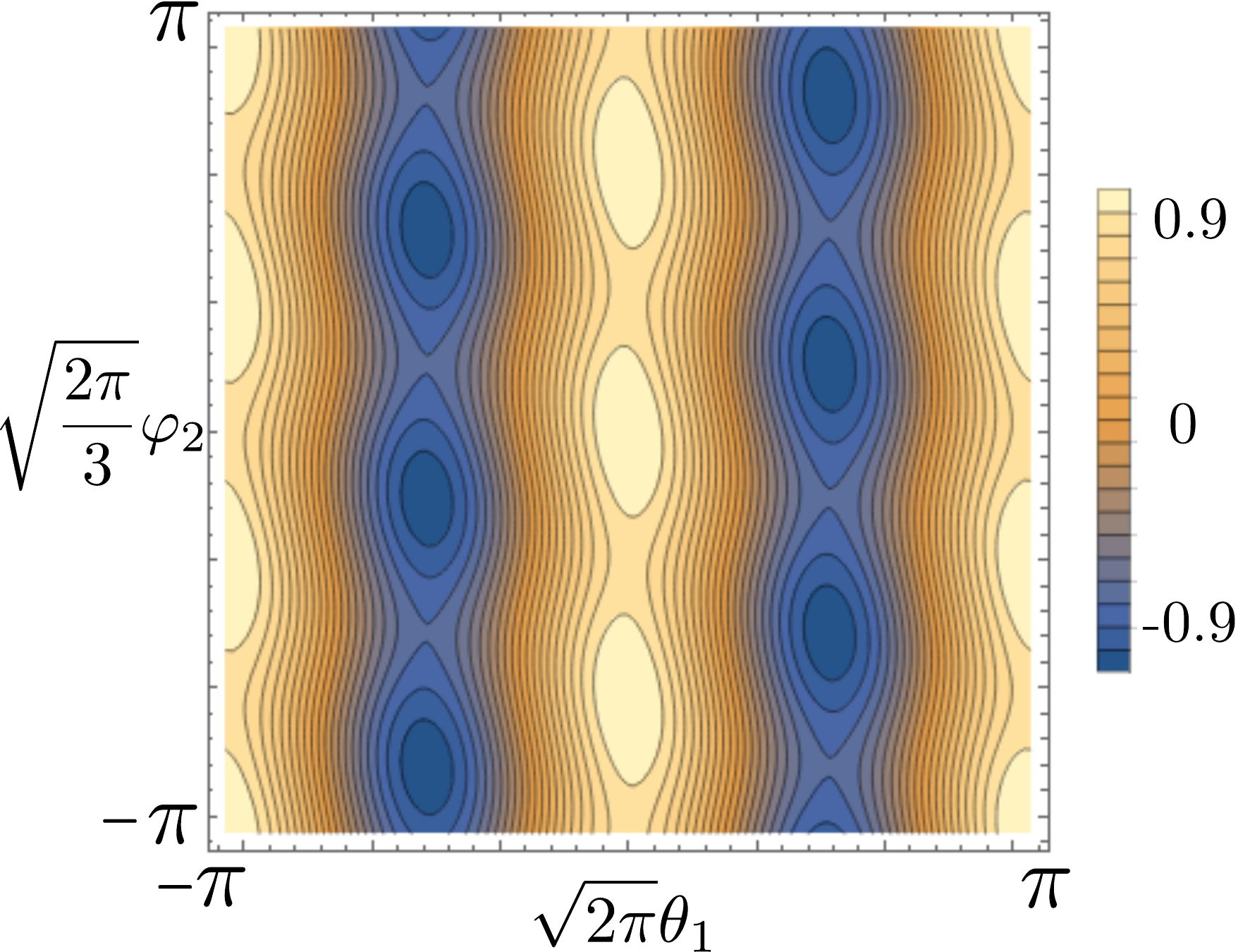}
\caption{The neutral part of %
$\mathcal{V}/|g_{\bar{3}}|$, 
for $g_1/g_{\bar{3}}=-0.15$, $g_2/g_{\bar{3}}=0.1$ and $g_{\bar{3}}>0$.
Taking the compactification of the fields into account, there are six inequivalent minima (blue). For a given charge $\tilde{N}_c$ and $\mathbb{Z}_3$ label $\tilde{N}_1 \text{ mod }3$, the ground state corresponds to a superposition involving all six minima.}\label{fig:potential2D}
\end{figure}

The neutral sector works analogously. 
We focus on the chiral clock model phenomenology arising for $g_{\bar{3}}>0$, a regime suggested by the contributions of strong repulsive total-density interactions.
In the absence of chirality ($g_1=0$), a gap arises for large $|g_2|\gg g_{\bar{3}}$; in this case the fields $(\sqrt{2\pi}\theta_1,\sqrt{6\pi}\varphi_2)$ are locked to the configuration
$\pi[n_x,n_x+2n_y+\Theta(g_2)]$.
For large $g_{\bar{3}}\gg |g_2|$, the system is gapless if $g_1=0$ and has central charge $c=1$. 
These features for $|g_2|\gg g_{\bar{3}}$ and $g_{\bar{3}}\gg |g_2|$ are similar to those of the ordered and incommensurate phases of the nonchiral regimes of the clock model, respectively. 
Furthermore, the chiral coupling $g_1\neq 0$ can open a gap for $g_{\bar{3}}\gg |g_2|$ (with locking configuration $\pi[n_x+\frac{1}{2},n_x+2n_y-\frac{1}{2}\text{sgn}(g_1)]$), while for $|g_2|\gg g_{\bar{3}}$ it can close the gap (provided $|g_1|\approx |g_2|$), similarly to the effects of chiral deformations on the incommensurate and ordered phases of the clock model. 
The neutral part of $\mathcal{V}$ for $g_{\bar{3}}\gg |g_2|$ and $g_1\neq 0$ is illustrated in Fig.~\ref{fig:potential2D}.

The correspondence between the fermions $\psi_{\eta\alpha}$ and the fields $\theta_\mu$, $\varphi_\mu$ implies that the latter are periodic variables. In particular, $(\theta_{c},\theta_{1},\varphi_{2})\sim(\theta_{c},\theta_{1},\varphi_{2})+\sqrt{\pi}\bm{n}\cdot\bm{\Gamma}$ (together with a suitable shift of the conjugate pairs) where $\bm{n}$ is a vector of integers and $\bm{\Gamma}=\left(\begin{smallmatrix}
\sqrt{3} & 0 & 0\\
0 & \sqrt{2} & 0\\
1/\sqrt{3} & -3/\sqrt{2} & 1/\sqrt{6}
\end{smallmatrix}\right)$. 
Under this compactification, the six minima in Fig.~\ref{fig:potential2D}, at a given $\theta_{c}$,  are inequivalent. Six inequivalent minima can also be arranged using a single $\theta_1$ (and $\theta_c$) but doubling the $\varphi_2$ interval. 
The operator $\exp(2\pi i\tilde{N}_{2}/6)$, which commutes with  $H_\text{b}$, toggles between the six minima in the latter arrangement; since the values of $\tilde{N}_{c,1}$ specify $\exp(2\pi i\tilde{N}_{2}/6)$, for a given $\tilde{N}_{c,1}$ there is a single ground state corresponding to a superposition between the six minima. 
The presence of $\mathbb{Z}_3$ symmetry, however, guarantees the conservation only of $\tilde{N}_{1}\text{ mod }3$; e.g., boundary terms (or corresponding neutral-sector boundary conditions) generically couple states with $\tilde{N}_1$ and $\tilde{N}_1+3$. For fixed particle number $\tilde{N}_c$, we are thus left with three towers of excitations, each labeled by its $\mathbb{Z}_3$ eigenvalue.

{\it Zero modes}.-
A $\mathbb{Z}_3$ label in itself does not imply degeneracies in the spectrum. 
However, a threefold degeneracy arises if a zero-mode operator $\chi$ exists that commutes with the Hamiltonian while toggling the $\mathbb{Z}_3$ label.
To obtain such zero modes, we consider the ground-state projections $\chi_{0,\ell}$ of neutral operators at the edge ($x=0,\ell$), taking operators that change $\tilde{N}_1 \text{ mod }3$. 
Neutrality ensures maintaining a fixed particle number, while being at the edge means that $\chi_{0,\ell}$ create in $(\theta_1,\varphi_2)$ only boundary kinks, a feature compatible~\cite{Lindner2012,Cheng2012,Clarke2013,Mong2014} with commutation with at least the groundstate (or, more generally, sub-gap) sector of the Hamiltonian. 
As in parafermion systems, there are several choices for $\chi_{0,\ell}$~\cite{Mong2014}.
We choose $\chi_{0,\ell}$  as the ground-state projection of an operator $\mathcal{O}_\tau(x=0,\ell)$  in the expansion of $f_{1}^{\dagger}f_{2}$, with
$\mathcal{O}_{\tau}^\dagger \propto e^{i\frac{1}{3}\sqrt{12\pi}\theta_{c}}e^{i\sqrt{2\pi}\varphi_{1}}e^{i\frac{1}{3}\sqrt{6\pi}\theta_{2}}$.
Hence, $\chi_{0,\ell}$ are local operators with  $\chi_0\chi_\ell=\chi_\ell \chi_0$. 
They also satisfy $\chi_{0,\ell}\mathcal{S}=\omega\mathcal{S}\chi_{0,\ell}$, the requisite toggling of the $\mathbb{Z}_3$ label. 
By introducing the nonlocal combination $\chi_{\ell}^{\prime}=\mathcal{S}\chi_{\ell}$, we can also use a pair of mutual $\mathbb{Z}_3$-parafermions: $\chi_{0}\chi_{\ell}^{\prime}=\omega\chi_{\ell}^{\prime}\chi_{0}$. 
The phenomenology recovered here matches that of the chiral clock model~\cite{Fendley2012,Fendley2014,Fendley2016,Alicea2016}, including the nonlocal nature of its parafermions. 
However bosonization indicates the zero-mode character of $\chi_{0,\ell}^{(\prime)}$ only for sub-gap energies, which is sufficient only for their status as ``weak'' zero modes~\cite{Alicea2016}. 
To study whether strong zero modes emerge, and for a concrete illustration of our bosonization phenomenology, we next turn to our microscopic model Eq.~\eqref{Ham_lattice}.

{\it Strong-interaction perturbation theory}.- 
We work deep in the Mott insulator regime, $U\gg |J|$. 
We start with $J=J_\perp=V_a=0$ in Eq.~\eqref{Ham_lattice}. 
This limit has a highly degenerate multiplet of lowest-energy states, each with one fermion per site.
Turning on a small $V_a$ and $J_\perp$ splits the low-energy states, without coupling to high-energy states with more than one fermion per site. 
In contrast, intraleg tunneling $H_J=-\frac{J}{2}\sum_{a,i}c_{i,a}^\dagger c_{i+1,a}+\text{H.c.}$ connects low- and high-energy states. 
To obtain the Hamiltonian governing the physics of low-energy states, we perform perturbation theory in $J/U$. 
The first nonzero correction is in second order in $J/U$, corresponding to a $\mathbb{Z}_3$ analogue of superexchange. We find the effective Hamiltonian
\begin{eqnarray}\nonumber
 H_{\rm eff}&=& \sum_{j,a,b}\left(V_b n_{j,a}n_{j+1,a+b}+\frac{J^2}{2U}c_{j,a}^{\dagger}c_{j+1,a}c_{j+1,b}^{\dagger}c_{j,b}\right)\\
 &+&J_\perp\sum_{j,a}(c_{j,a}^{\dagger}c_{j,a+1}+\text{H.c.}),\label{H_eff_fermion}
\end{eqnarray}
valid for $U\gg |V_a|,|J|,|J_\perp|$.

Using the single occupation per site constraint, $\sum_a n_{j,a}=1$, we can rewrite $H_{\rm eff}$ in a form where the relation to the chiral clock model becomes manifest~\cite{supplementary}. 
We map fermion bilinears to matrices 
$c^\dagger_{j,a}c_{j,b}\rightarrow M_j^{ab}$, with components $(M_j^{ab})_{kl}=\delta_{ak}\delta_{bl}$ (omitting factors of identity away from site $j$). We find
\begin{eqnarray}\label{H_eff_matrix}
\!\!\!\!\!\!H_{\rm eff}= \!\sum_{j=1}^N J_\perp\sigma_j+\!\!\sum_{j=1}^{N-1}\frac{V'e^{i\Phi}}{2}\tau_j\tau^\dagger_{j+1}\!+\!\frac{J^2}{4U}P_{j,j+1}\!+\!\text{H.c.},
\end{eqnarray}
where we have used Eq.~\eqref{parameters_chiral} and $N=\ell/a_0$. Here $\tau_j$ and $\sigma_j$ are the clock variables
\begin{equation}
 \tau_j=\left(\begin{array}{ccc}
 1&0&0\\
 0&\omega&0\\
 0&0&\omega^2
 \end{array}\right),\quad 
  \sigma_j=\left(\begin{array}{ccc}
 0&0&1\\
 1&0&0\\
 0&1&0
 \end{array}\right),
\end{equation}
while $P_{j,j+1}|a\rangle_{j}|b\rangle_{j+1}  =|b\rangle_{j}|a\rangle_{j+1}$ is the swap operator between neighboring sites.
For $J=0$,  Eq.~\eqref{H_eff_matrix} recovers the three-state quantum clock model. 
It is chiral for $\Phi\neq 0 \text{ mod }\pi/3$~\cite{Fendley2012,Alicea2016}. Eqs.~\eqref{parameters_chiral} and~\eqref{H_eff_matrix} thus explicitly establish the link between geometric and clock-model chirality.  
For $J\neq0$, the clock model is perturbed by the Uimin-Lai-Sutherland Hamiltonian~\cite{Uimin1970,Lai1974, Sutherland1975}, seen to arise here from superexchange. 

\begin{figure}
 \includegraphics[width=0.95\columnwidth]{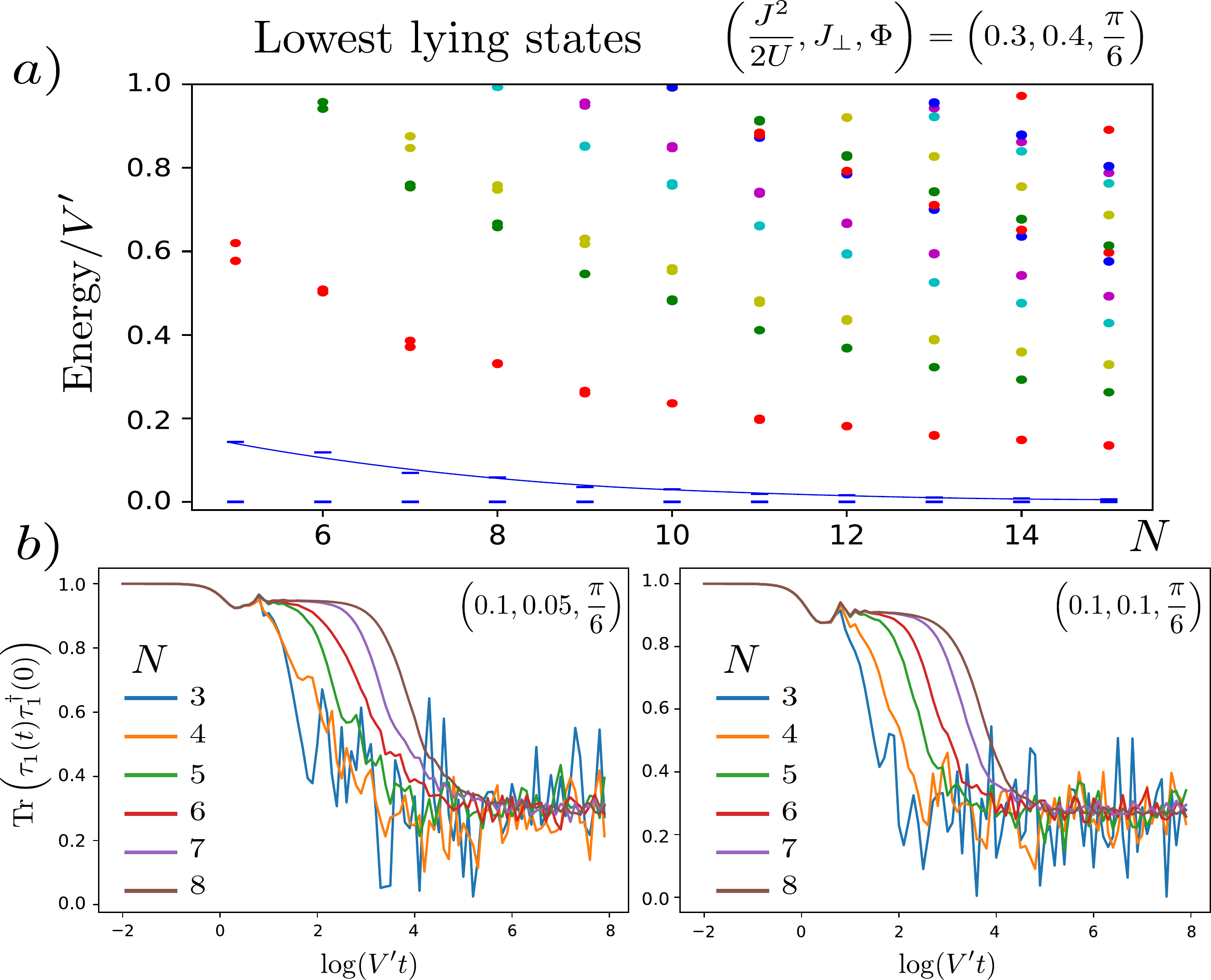}
\caption{a.- Low-lying energies of $H_\text{eff}$, measured from the ground state, for various system sizes $N$. 
The coloring changes for every three consecutive levels, repeating after the sixth color.
The values of $J^2/2U$ and $J_\perp$ (shown at the top, with $\Phi$) are measured in units of $V^\prime$. 
The lowest excited states have energy (blue guide for the eye) consistent with ground-state-triplet splitting decaying exponentially with $N$. 
The excited-state-triplet splittings are barely visible on the scale of the figure.
b.- The dynamical correlator ${\rm Tr}(\tau_1(t)\tau^\dagger_1(0))$. The persistence close to the initial value for times exponentially long in $N$ indicates the presence of strong zero modes.}\label{fig:zero_mode}
\end{figure}

For $J=0$, weak $J_\perp$, and $\Phi$ sufficiently away from $0 \text{ mod } \pi/3$, the system supports  strong zero modes with the same properties as that of $\chi_{0,\ell}^{(\prime)}$ above~\cite{Fendley2012}. 
Our next goal is to assess whether these strong zero modes survive the superexchange perturbation.
We expect this for $|V^\prime \sin(3\Phi)|\gg |J_\perp|,J^2/U$: here the combination of the spectral separation between various \mbox{$J=J_\perp=0$} domain-wall sectors and the scale separation between the chirality-induced intra-sector splittings versus small $|J_\perp|,J^2/U$ might offer protection also against small nonzero $J$ thanks to the restricted manner in which the swap operation acts on domain-wall states~\cite{supplementary}. 
To assess the presence of strong zero modes, we first perform exact diagonalization to study the energy spectrum in this regime
(Fig. \ref{fig:zero_mode}a). 
The existence of a zero mode that arranges the spectrum in $\mathbb{Z}_3$ triplets is manifest, both for the three lowest-lying states whose energy splitting decays consistently with an exponential in $N$, as well as for higher energies where the much smaller triplet splitting is almost invisible on the scale of the figure.
Our findings are further corroborated by the $5<N<100$ excited-state spectrum in the single-domain-wall sector, where
the $\mathbb{Z}_3$-triplet splittings decay exponentially with $N$  to a value numerically indistinguishable from zero~\cite{supplementary}.

We next study dynamical consequences of these zero modes. 
Their edge-mode nature and exchange properties with $\mathcal{S}=\prod_{j=1}^N \sigma_j$ suggest that they contribute significantly to $\tau_j$ at the boundary (in fact, $\tau_{1,N}$ 
are commuting zero modes for $J=J_\perp=0$~\cite{Fendley2012,Alicea2016}, akin to $\chi_{0,\ell}$). 
We compute the infinite-temperature correlator~\cite{Kemp2017,Else2017}
\begin{equation}
{\rm Tr}(\tau_1(t)\tau^\dagger_1(0))=3^{-N}\sum_{k,j}|\langle k|\tau_1|j\rangle|^2e^{i(E_k-E_j)t},
\end{equation}
where $E_{j}$ is the energy of the eigenstate $|j\rangle$.
Time-independent contributions to the correlator come only from $|k\rangle$, $|j\rangle$ in degenerate triplets, provided $\langle k|\tau_1|j\rangle\neq 0$.
(Diagonal matrix elements vanish due to $\tau_1 \mathcal{S} = \omega  \mathcal{S} \tau_1$.)
For the signal to be appreciable, $\langle k|\tau_1|j\rangle\neq 0$ should hold for many degenerate triplets. 
In Fig. \ref{fig:zero_mode}b we show the numerically evaluated correlator for two values of the parameters. (The behavior is similar for other values in the $|V^\prime \sin(3\Phi)|\gg |J_\perp|,J^2/U$ regime, even for weakly disordered systems~\cite{supplementary}.)
The results are consistent with the correlator remaining close to its initial value for times exponentially long in system size.  
Such behavior indicates exponentially decaying triplet splittings across the whole spectrum,
and thus illustrates how strong zero modes lead to long-time coherence far from the ground state. 

{\it Conclusions}.-
We have shown how geometric chirality in a $\mathbb{Z}_3$- and $\mathcal{T}$-invariant Mott insulator can lead to chiral quantum clock model physics, including strong zero modes. 
We illuminated this from two complementary perspectives: bosonization based on general symmetry considerations, and the analysis of the model Eq.~\eqref{Ham_lattice} for strong interactions $U\gg |J|$. 
In our bosonization, after implementing the Mott-insulating regime by gapping out the charge degrees of freedom, 
we 
uncovered fingerprints of the ordered and incommensurate phases of the chiral quantum clock model, and established the presence of zero modes in the ordered phase.  
This suggests that chiral interactions %
provide a favorable setting for chiral-quantum-clock-model physics to arise 
in Mott insulators.
Our model Eq.~\eqref{Ham_lattice} gives concrete evidence for this:
for $U\gg |J|$, we explicitly recover the chiral quantum clock model, perturbed by the Uimin-Lai-Sutherland Hamiltonian via superexchange. 
Our simulations suggest that the strong zero modes survive the superexchange perturbation, leading to the correlators of clock operators at the edge to persist near their initial value for times exponentially long in system size, even at infinite temperature.
Interesting directions for future research include the study of the phase diagram and dynamics of our model  with approaches that  probe the thermodynamic limit. This will, in particular, illuminate where the zero modes we found fall in the refined ``strong" versus ``almost strong" classification of Refs.~\cite{Kemp2017,Else2017}.

We acknowledge discussions with D. Gutman, K. Snizhko, F. Buccheri, and J. Park, and we thank P. Fendley for feedback on the manuscript. This work was supported by the ERC Starting Grant No. 678795 TopInSy.

\pagebreak

\onecolumngrid
\section*{Supplemental Material}

In this Supplemental Material we explain the correspondence between Eq.~(4) and (5) of the main text, and provide further corroborating evidence complementing that in Fig. 4 of the main text.

\section{From the fermion Hamiltonian to the perturbed quantum clock model}
\label{sec:map_fer}
In the main text we note that the constraint of single occupation per site, $\sum_an_{j,a}=1$,  allows us to rewrite the fermion Hamiltonian [Eq. (4) of the main text] in the form of a perturbed chiral quantum clock model [Eq. (5) of the main text] using a map between fermion bilinears and matrices $c_{j,a}^\dagger c_{j,b} \rightarrow M_{j}^{ab}$. In this section, we provide the details behind these observations. 

The matrix $M_{j}^{ab}$ is understood as
\begin{equation}
M_{j}^{ab}=\openone_1\otimes\openone_2\dots\openone_{j-1}\otimes M^{ab} \otimes \openone_{j+1}\dots\openone_{N},
\label{eq:ferm_bilinears}
\end{equation}
where the entries of the $3\times3$ matrix $M^{ab}$ are $(M^{ab})_{kl}=\delta_{ak}\delta_{bl}$. 
The mapping $c_{j,a}^\dagger c_{j,b} \rightarrow M_{j}^{ab}$ is possible because the single-occupancy condition sets the local Hilbert space dimension ${\rm dim}(\mathcal{H}_j)=3$ for each site $j$. 
We define a basis of this local Hilbert space as
$|l\rangle_j=c^\dagger_{j,l}|v\rangle_j$, where $|v\rangle_j$ is the state with no fermions at site $j$.
Every operator $\mathcal{O}$ that preserves the single occupancy per site condition can be written in this basis using 
\begin{equation}
 \mathcal{O} = \sum_{l_1\dots l_N=0}^2\sum_{k_1\dots k_N=0}^2\left(\bigotimes_{j=1}^N|k_j\rangle\langle k_j|\right)\mathcal{O}\left(\bigotimes_{i=1}^N|l_i\rangle\langle l_i|\right),
\end{equation}
where we have introduced the resolution of the identity $\sum_{l_j=0}^2|l_j\rangle\langle l_j|=\openone_j$ at each lattice position. 
For example, for an operator $\mathcal{A}_{i}\mathcal{B}_j$ that acts nontrivially at sites $i$ and $j$ ($i<j$), we have
\begin{equation}
\mathcal{A}_{i}\mathcal{B}_j=\sum_{l_i,l_jk_ik_j=0}^2\openone_1\otimes\dots\openone_{i-1}\otimes |k_i\rangle\langle k_i|\mathcal{A}_i|l_i\rangle\langle l_i|\otimes\openone_{i+1}\otimes\dots|k_j\rangle\langle k_j|\mathcal{B}_j|l_j \rangle\langle l_j|\otimes \openone_{j+1}\dots\openone_{N}.
\end{equation}
This provides a complete definition of $\mathcal{A}_{i}\mathcal{B}_j$ in terms of the matrix elements in each local Hilbert space.
For the bilinear $c^\dagger_{j,a}c_{j,b}$, the matrix elements are
\begin{equation}
 (M^{ab})_{kl}=\langle k_j|c^\dagger_{j,a}c_{j,b}|l_j\rangle=\langle v|_jc_{j,k}c_{j,a}^\dagger c_{j,b}c^\dagger_{j,l}|v\rangle_j=\delta_{ak}\delta_{bl}.
\end{equation}
The relation between the $c_{j,a}^\dagger c_{j,b} \rightarrow M_j^{ab}$ map and the basis used for the matrices [Eq. (6) of the main text]
\begin{equation}
 \tau_j=\left(\begin{array}{ccc}
 1&0&0\\
 0&\omega&0\\
 0&0&\omega^2
 \end{array}\right),\quad 
  \sigma_j=\left(\begin{array}{ccc}
 0&0&1\\
 1&0&0\\
 0&1&0
 \end{array}\right)
\end{equation}
corresponds to the assignments 
\begin{equation}\label{states}
 c^\dagger_{j,0}|v\rangle_j=\begin{bmatrix}1\\
0\\
0
\end{bmatrix}_j,\quad
c^\dagger_{j,1}|v\rangle_j=\begin{bmatrix}0\\
1\\
0
\end{bmatrix}_j,\quad\mbox{and}\quad c^\dagger_{j,2}|v\rangle_j=\begin{bmatrix}0\\
0\\
1
\end{bmatrix}_j.
\end{equation}

To get from the fermion Hamiltonian [Eq. (4) of the main text] to the perturbed chiral quantum clock model [Eq. (5) of the main text], we invert the relations
\begin{equation}\label{parameters_chiral_app}
V=\frac{1}{3}\sum_aV_{a},\qquad V'\sin\Phi =\frac{V_{1}-V_{2}}{\sqrt{3}},\qquad V'\cos\Phi =\frac{2V_{0}-V_{1}-V_{2}}{3}
\end{equation}
to find 
\begin{equation}
V_0=V+V^{\prime}\cos\Phi,\qquad V_1=V+V^{\prime}\cos\left(\Phi-\frac{2\pi}{3}\right),\qquad V_2=V+V^{\prime}\cos\left(\Phi+\frac{2\pi}{3}\right). 
\end{equation}
Combining this, the constraint $\sum_an_{j,a}=1$, and the mapping Eq.~\eqref{eq:ferm_bilinears}, we find that, up to a constant offset, Eq. (4) of the main text maps to the Hamiltonian in Eq. (5) of the main text, 
\begin{equation}\label{eq:Heff_5}
H_{\rm eff}= \sum_{j=1}^N J_\perp\sigma_j+\sum_{j=1}^{N-1}\left(\frac{V'e^{i\Phi}}{2}\tau_j\tau^\dagger_{j+1}+\frac{J^2}{4U}P_{j,j+1}\right)+\text{H.c.},
\end{equation}
repeated here for convenience.

\section{Stability of the strong zero modes}
\label{sec:zero_mode}

The strong zero modes are expected to be present for $|V^\prime \sin(3\Phi)|\gg |J_\perp|,J^2/U$ (cf also Sec.~\ref{sec:1dw}). 
We investigate this by studying the system-size dependence of (i) the splittings within $\mathbb{Z}_3$ triplets and (ii) the dynamics of the correlator ${\rm Tr}(\tau_1(t)\tau^\dagger_1(0))$. 
Using exact diagonalization of Eq.~\eqref{eq:Heff_5} [Eq. (5) of the main text], we show that the system continues to display the behavior exemplified in Fig. 4 of the main text for a range of parameters, and even in the presence of weak disorder. 
We shall also find, by investigating the excited-state spectrum in the single-domain-wall sector for system sizes far beyond that achievable by exact diagonalization, that in the entire range of the $\mathbb{Z}_3$-triplet splittings that we can resolve by machine precision, the splittings continue to decay exponentially with system size for a range of parameters, thus indicating the presence of strong zero modes.

\subsection{$\mathbb{Z}_3$-triplet splittings}
\label{sec:spectral-signatures}

\subsubsection{Exact diagonalization}
\begin{figure}%
\includegraphics[scale=0.45]{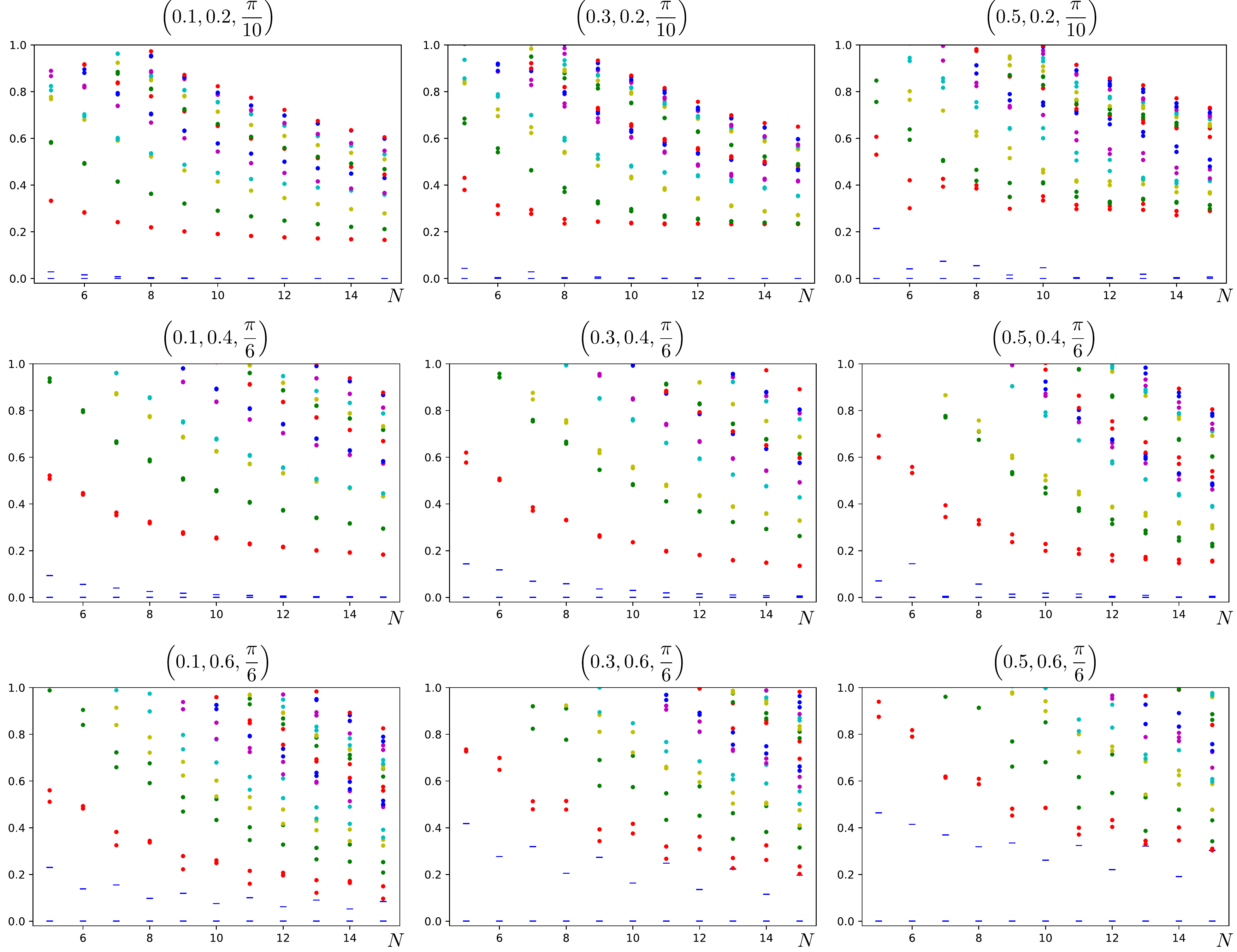}
    \caption{Low-lying spectrum as function of the system size $N$ for different parameters $\left(\frac{J^2}{2U},J_\perp,\Phi\right)$, obtained by exact diagonalization of Eq. (5) of the main text. 
    The energies are measured in units of $V^\prime$, and are relative to the ground state. 
    The ground state and the two lowest levels are shown by blue dashes.
    Every three next levels (i.e., each $\mathbb{Z}_3$ triplet) are depicted by the same color (the colors repeat after six consecutive triplets). 
    For $J_\perp\leq 0.4$, we see that the levels within a triplet approach each other as the system size increases. For $J_\perp=0.6$, the results are inconclusive, but point to a destruction of the degeneracies and hence of the strong zero modes.}
 \label{fig:low_lying_uni}
\end{figure}

We first study the low-lying spectrum of Eq.~\eqref{eq:Heff_5}  using numerical exact diagonalization. 
In Fig.~\ref{fig:low_lying_uni} we show results for the clean system (i.e., without disorder) for various values of the parameters. 
We display energy levels belonging to a $\mathbb{Z}_3$ triplet using the same color (the colors repeat after the the sixth consecutive triplet).
The main parameter influencing how the $\mathbb{Z}_3$-triplet splittings decay with system size $N$ is observed to be $J_\perp/V'$. 
For small values of $J_\perp/V'$, the behavior is consistent with the presence of strong zero modes: the triplet splittings  decay quickly with $N$. 
(In Sec.~\ref{sec:1dw}, we shall show that the decay is exponential in $N$.)
For larger values ($J_\perp=0.6V'$), the rapid decay with $N$ ceases, signaling a destruction of the zero modes.
That the parameter $J^2/2UV'$ is observed to have only minor effect is significant: it is known that for $J=0$ and small $J_\perp$ the system supports strong zero modes and our findings are consistent with the $J^2/2U$ perturbation preserving these. 

\begin{figure}%
\includegraphics[scale=0.45]{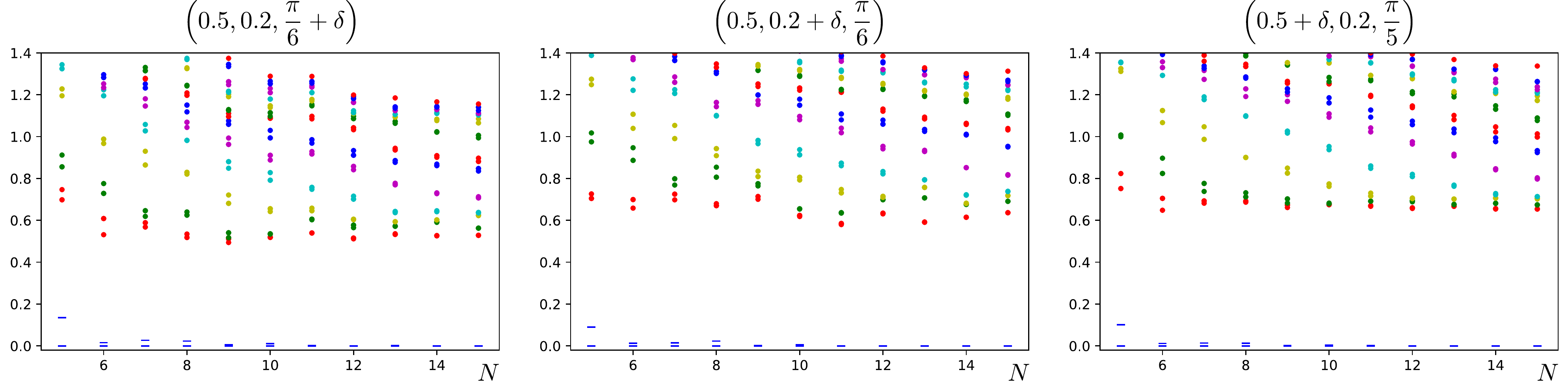}
    \caption{Low-lying spectrum for systems with weak disorder, as function of the system size $N$. 
    We disorder the parameters $\left(\frac{J^2}{2U},J_\perp,\Phi\right)$ using the random variable $\delta$, which for each site (bond) is uniformly distributed on the interval $[-0.1,0.1]$ with no spatial correlations. 
    The conventions are as in Fig.~\ref{fig:low_lying_uni}.
    We observe that the splittings within a triplet decay rapidly with $N$.
    }
 \label{fig:low_lying_rand}
\end{figure}

In Fig.~\ref{fig:low_lying_rand} we show results for systems with three forms of weak, spatially uncorrelated, disorder. 
The observed behavior is consistent with the presence of strong zero modes.

\pagebreak
\subsubsection{Single-domain-wall spectrum}
\label{sec:1dw}

To quantify the decay of the excited-state $\mathbb{Z}_3$-triplet splittings with the system size $N$, we must reach much larger values of $N$ than those permitted by the exponential-in-$N$ scaling problem size in exact diagonalization. 
To this end, we focus on phases $\Phi$ in the vicinity of $\Phi=\pi/3$, where an energy window above the ground state, with an extensive number of levels,  can be captured by an effective Hamiltonian of size $\propto N$,  acting in the subspace of the Hilbert space spanned by the  single-domain-wall excitations of 
\begin{equation}\label{HeffJJ0}
H_{\rm eff}(J=J_\perp=0)=\frac{V'}{2}\sum_{i}\left(\tau_{j}\tau_{j+1}^{\dagger}e^{i\Phi}+\tau_{j}^{\dagger}\tau_{j+1}e^{-i\Phi}\right).
\end{equation}

Our approach generalizes that of Ref.~\cite{Jermyn2014} by the inclusion of the Uimin-Lai-Sutherland perturbation. 
To make contact with this work, we note that the parameters $V^\prime$, $\Phi$ and $J_\perp$ in our $H_{\rm eff}$ [Eq. (5) of the main text and Eq.~\eqref{eq:Heff_5} above] and those of Eq. (9) in Ref.~\cite{Jermyn2014} (indicated here by the subscript ``ref")  are related as follows:
\begin{equation}
V'=2J_{\rm ref},\quad \Phi = \phi_{\rm ref}+\pi,\quad J_\perp = -f_{\rm ref}.
\end{equation}
Owing to the unitary equivalence (see e.g., Ref.~\cite{Zhuang2015}) relating $\Phi$ and $\Phi+2\pi n/3$ (with $n$ integer), near $\Phi=\pi/3$ and for $J=0$ our model shares the same physics as that of Ref.~\cite{Jermyn2014} for $\phi_{\rm ref}=0$. 
It is near this ``nearly ferromagnetic" regime where an effectively noninteracting (hence scaling as $\sim N$) description in terms of the single-domain-wall sector is possible. (Here, ``nearly ferromagnetic" is understood in the sense of Ref.~\cite{Jermyn2014}; under the unitary transformation relating their model at $\phi_{\rm ref}=0$ to ours at $\Phi=\pi/3$, the ferromagnetic $\phi_{\rm ref}=0$ ground states of Ref.~\cite{Jermyn2014} are mapped to the $\Phi=\pi/3$ ground states of our model which have ``antiferromagnetic" or ``spiral" character. However, to avoid confusion, we do \emph{not} call our $\Phi=\pi/3$  ``antiferromagnetic" point, but follow the parlance of Ref.~\cite{Jermyn2014}. In this sense, the antiferromagnetic point in our model is at $\Phi=0$.)

To obtain the single-domain-wall Hamiltonian $H_\text{1dw}$, we perform first-order pertubation theory in $J^2/2U$ and $J_\perp$, for the Hamiltonian Eq.~\eqref{eq:Heff_5}.
For $J=J_\perp=0$ and $\Phi\approx\frac{\pi}{3}$, the ground states are $|{\bf A}\rangle\equiv|021021...\rangle,|{\bf B}\rangle\equiv|102102...\rangle$, and $|{\bf C}\rangle\equiv|210210...\rangle$, where the entry $a=0,1,2$ at site $j$ refers to $|a\rangle_j$ understood as the right-hand-sides of Eq.~\eqref{states}. (Tensor product signs are omitted in $|{\bf A}\rangle$, $|{\bf B}\rangle$, $|{\bf C}\rangle$.)
The ground state energy is $E_{0}=V'(N-1)\cos(\Phi+\frac{2\pi}{3})$.
The states $|a\rangle_j$ transform under the $\mathbb{Z}_3$ symmetry $\mathcal{S}$ as $\mathcal{S}|a\rangle	=|a+1\rangle$, $(a+3\equiv a)$.

There are six types of lowest excited states, given by the domain wall configurations formed from the three previous ground states. 
Explicitly, the domain wall configurations are
\begin{eqnarray}
|{\bf A}|_{i}{\bf B}\rangle	=\prod_{j=1}^{i}\sigma_{j}^{\dagger}|{\bf B}\rangle,\quad |{\bf B}|_{i}{\bf C}\rangle	=\prod_{j=1}^{i}\sigma_{j}^{\dagger}|{\bf C}\rangle,\quad |{\bf C}|_{i}{\bf A}\rangle	=\prod_{j=1}^{i}\sigma_{j}^{\dagger}|{\bf A}\rangle,\\
|{\bf A}|_{i}{\bf C}\rangle	=\prod_{j=1}^{i}\sigma_{j}|{\bf C}\rangle,\quad
|{\bf B}|_{i}{\bf A}\rangle	=\prod_{j=1}^{i}\sigma_{j}|{\bf A}\rangle,\quad
|{\bf C}|_{i}{\bf B}\rangle	=\prod_{j=1}^{i}\sigma_{j}|{\bf B}\rangle,
\end{eqnarray}
with $i \in[1, N-1]$. 
They are related by the $\mathbb{Z}_{3}$ symmetry $\mathcal{S}$ as 
\begin{eqnarray}
 \mathcal{S}|{\bf A}|{\bf B}\rangle	=|{\bf B}|{\bf C}\rangle,\quad 
 \mathcal{S}|{\bf B}|{\bf C}\rangle	=|{\bf C}|{\bf A}\rangle,\quad 
 \mathcal{S}|{\bf C}|{\bf A}\rangle	=|{\bf A}|{\bf B}\rangle,\\
 \mathcal{S}|{\bf A}|{\bf C}\rangle	=|{\bf B}|{\bf A}\rangle,\quad
 \mathcal{S}|{\bf B}|{\bf A}\rangle	=|{\bf C}|{\bf B}\rangle,\quad
 \mathcal{S}|{\bf C}|{\bf B}\rangle	=|{\bf A}|{\bf C}\rangle.
\end{eqnarray}
The symmetry therefore subdivides the set of single domain wall states into two conjugacy classes $\{|{\bf A}|{\bf B}\rangle,|{\bf B}|{\bf C}\rangle,|{\bf C}|{\bf A}\rangle\}$ and $\{|{\bf A}|{\bf C}\rangle,|{\bf B}|{\bf A}\rangle,|{\bf C}|{\bf B}\rangle\}$ that transform into themselves under the symmetry $\mathcal{S}$. 

The two conjugacy classes of domain walls involve the local configurations $|00\rangle$,$|11\rangle$,$|22\rangle$ and $|01\rangle$,$|12\rangle$,$|20\rangle$, respectively, that are penalized by the Hamiltonian. The corresponding energies are $V'\cos\Phi$ and $V'\cos(\Phi-\frac{2\pi}{3})$ above the groundstates, respectively.
This dependence on $\Phi$ highlights the key feature brought in by chirality, i.e., as we move away from $\Phi=\pi/3$ (note that for $\Phi=\pi/3$ the physics is nonchiral, owing to the unitary transformation sending $\Phi\rightarrow \Phi+2\pi/3$): the two conjugacy classes are split in energy by an amount $\sim V^\prime$. 
This splitting, when it is much larger than the energy scales $J^2/2U$ and $J_\perp$ of the perturbations, introduces an energy barrier which, together with the restricted manner in which the perturbations act on domain-wall states, will be seen to be responsible for the exponentially decaying $\mathbb{Z}_3$-triplet splitting indicative of the strong zero mode~\cite{Fendley2012,Jermyn2014,Alicea2016}.

For $J_\perp\ll V'$ and $\frac{J^2}{2U}\ll V'$, degenerate first order perturbation theory mixes the single-domain-wall states. 
We first discuss the effect of the $J_\perp$ perturbation~\cite{Jermyn2014}.
In the bulk of the system (away from the boundaries), the Hamiltonian $H_{J_\perp}=J_\perp\sum_{i=2}^{N-1}(\sigma_i^\dagger+\sigma_i)$ induces hopping of a domain wall, and hence leads to a bandwidth $\sim J_\perp$~\cite{Jermyn2014}. 
On the left boundary, the action of $J_{\perp}(\sigma_1+\sigma_1^\dagger)$ within the single-domain-wall subspace is
\begin{equation}
|{\bf B}|_{1}{\bf A}\rangle\leftrightarrow|{\bf C}|_{1}{\bf A}\rangle,\quad
|{\bf A}|_{1}{\bf C}\rangle\leftrightarrow|{\bf B}|_{1}{\bf C}\rangle,\quad
|{\bf C}|_{1}{\bf B}\rangle\leftrightarrow|{\bf A}|_{1}{\bf B}\rangle.
\end{equation}
On the right boundary, $J_{\perp}(\sigma_N+\sigma_N^\dagger)$ generates the processes
\begin{equation}
|{\bf B}|_{N-1}{\bf A}\rangle\leftrightarrow|{\bf B}|_{N-1}{\bf C}\rangle,\quad
|{\bf A}|_{N-1}{\bf C}\rangle\leftrightarrow|{\bf A}|_{N-1}{\bf B}\rangle,\quad
|{\bf C}|_{N-1}{\bf B}\rangle\leftrightarrow|{\bf C}|_{N-1}{\bf A}\rangle.
\end{equation}

The perturbation $H_{\rm S}=\frac{J^2}{2U}\sum_{i}P_{i,i+1}$ can also be analyzed using first order perturbation theory. 
In the bulk, the only nontrivial action of $H_{\rm S}$ preserving the single-domain-wall sector in the conjugacy class $\{|{\bf A}|{\bf B}\rangle,|{\bf B}|{\bf C}\rangle,|{\bf C}|{\bf A}\rangle\}$; here the swap invariance of the local configurations $|00\rangle$,$|11\rangle$,$|22\rangle$ results in a mere shift of the energy by $J^2/2U$. (Note that the absence of a nontrivial action for the complementary class means that, similarly to chirality, $H_{\rm S}$ contributes to the splitting between conjugacy classes.)
At the left end of the system, we find that in the single-domain-wall subspace $H_{\rm S}$ generates the processes
\begin{equation}
 P_{12}|{\bf A}|_{1}{\bf C}\rangle=|{\bf B}|_{2}{\bf C}\rangle,\quad 
 P_{12}|{\bf B}|_{1}{\bf A}\rangle=|{\bf C}|_{2}{\bf A}\rangle,\quad
 P_{12}|{\bf C}|_{1}{\bf B}\rangle=|{\bf A}|_{2}{\bf B}\rangle,
\end{equation}
together with the reverse processes
\begin{equation}
 P_{12}|{\bf B}|_{2}{\bf C}\rangle=|{\bf A}|_{1}{\bf C}\rangle,\quad 
 P_{12}|{\bf C}|_{2}{\bf A}\rangle=|{\bf B}|_{1}{\bf A}\rangle,\quad
 P_{12}|{\bf A}|_{2}{\bf B}\rangle=|{\bf C}|_{1}{\bf B}\rangle.
\end{equation}
On the right end of the chain, $H_{\rm S}$ generates the processes
\begin{eqnarray}
P_{N-1,N}|{\bf B}|_{N-1}{\bf A}\rangle=|{\bf B}|_{N-2}{\bf C}\rangle,\quad
P_{N-1,N}|{\bf A}|_{N-1}{\bf C}\rangle=|{\bf A}|_{N-2}{\bf B}\rangle,\quad
P_{N-1,N}|{\bf C}|_{N-1}{\bf B}\rangle=|{\bf C}|_{N-2}{\bf A}\rangle,\\
P_{N-1,N}|{\bf B}|_{N-2}{\bf C}\rangle=|{\bf B}|_{N-1}{\bf A}\rangle,\quad
P_{N-1,N}|{\bf A}|_{N-2}{\bf B}\rangle=|{\bf A}|_{N-1}{\bf C}\rangle,\quad
P_{N-1,N}|{\bf C}|_{N-2}{\bf A}\rangle=|{\bf C}|_{N-1}{\bf B}\rangle.
\end{eqnarray}

We therefore find, by combining the above results, that in the basis $\begin{bmatrix}v_{1} & v_{2} & v_{3} & \cdots & v_{N-3} & v_{N-2} & v_{N-1}\end{bmatrix}$, of the single-domain-wall subspace, where each $v_i$  represents 
\begin{equation}
 v_i = \begin{bmatrix}|{\bf A}|_{i}{\bf B}\rangle & |{\bf A}|_{i}{\bf C}\rangle & |{\bf B}|_{i}{\bf C}\rangle & |{\bf B}|_{i}{\bf A}\rangle & |{\bf C}|_{i}{\bf A}\rangle & |{\bf C}|_{i}{\bf B}\rangle\end{bmatrix},
\end{equation}
the single-domain-wall Hamiltonian $H_\text{1dw}$ is
\begin{equation}
H_\text{1dw}= \begin{bmatrix}L+D & Q+J_{\perp}I\\
Q^{\dagger}+J_{\perp}I & D & J_{\perp}I\\
 & J_{\perp}I & D & \ddots & J_{\perp}I\\
 &  & \ddots & \ddots & J_{\perp}I\\
 &  &  & J_{\perp}I & D & J_{\perp}I\\
 &  &  &  & J_{\perp}I & D & T+J_{\perp}I\\
 &  &  &  &  & T^{\dagger}+J_{\perp}I & D+R
\end{bmatrix}.
\end{equation}
Here, $I$ is the $6\times 6$ identity matrix,  
\begin{equation}
D=\begin{bmatrix}
\alpha & 0 & 0 & 0 & 0 & 0\\
0 & \beta & 0 & 0 & 0 & 0\\
0 & 0 & \alpha & 0 & 0 & 0\\
0 & 0 & 0 & \beta & 0 & 0\\
0 & 0 & 0 & 0 & \alpha & 0\\
0 & 0 & 0 & 0 & 0 & \beta
\end{bmatrix} \quad \mbox{where}\quad \alpha =\frac{J^2}{2U}+V'\cos\Phi \quad \mbox{and}  \quad \beta = V'\cos(\Phi-\frac{2\pi}{3}), 
\end{equation}
and
\begin{equation}
\quad L=J_{\perp}\begin{bmatrix}0 & 0 & 0 & 0 & 0 & 1\\
0 & 0 & 1 & 0 & 0 & 0\\
0 & 1 & 0 & 0 & 0 & 0\\
0 & 0 & 0 & 0 & 1 & 0\\
0 & 0 & 0 & 1 & 0 & 0\\
1 & 0 & 0 & 0 & 0 & 0
\end{bmatrix},\quad R=J_{\perp}\begin{bmatrix}0 & 1 & 0 & 0 & 0 & 0\\
1 & 0 & 0 & 0 & 0 & 0\\
0 & 0 & 0 & 1 & 0 & 0\\
0 & 0 & 1 & 0 & 0 & 0\\
0 & 0 & 0 & 0 & 0 & 1\\
0 & 0 & 0 & 0 & 1 & 0
\end{bmatrix},\quad Q=\frac{J^{2}}{2U}\begin{bmatrix}0 & 0 & 0 & 0 & 0 & 1\\
0 & 0 & 0 & 0 & 0 & 0\\
0 & 1 & 0 & 0 & 0 & 0\\
0 & 0 & 0 & 0 & 0 & 0\\
0 & 0 & 0 & 1 & 0 & 0\\
0 & 0 & 0 & 0 & 0 & 0
\end{bmatrix},\quad
T=\frac{J^{2}}{2U}\begin{bmatrix}0 & 1 & 0 & 0 & 0 & 0\\
0 & 0 & 0 & 0 & 0 & 0\\
0 & 0 & 0 & 1 & 0 & 0\\
0 & 0 & 0 & 0 & 0 & 0\\
0 & 0 & 0 & 0 & 0 & 1\\
0 & 0 & 0 & 0 & 0 & 0
\end{bmatrix}.
\end{equation}

\begin{figure}
\includegraphics[scale=0.51]{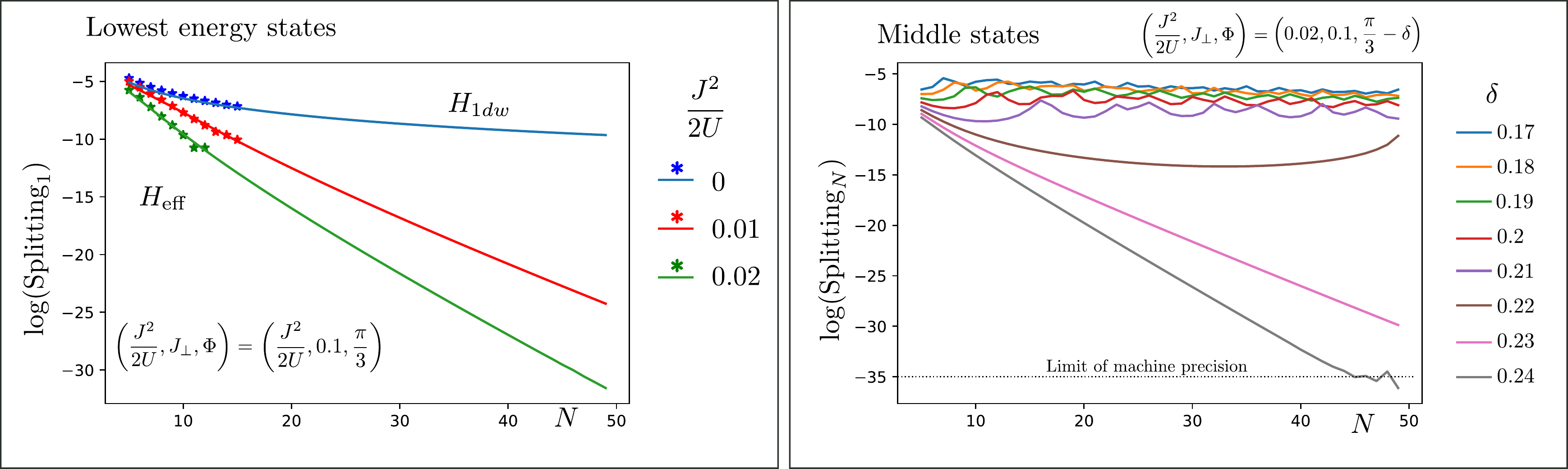}
    \caption{Splitting of energy levels of the single-domain-wall Hamiltonian $H_\text{1dw}$. Left: Splitting of the three lowest states as a function of the system size $N$. For small system sizes ($N\leq 15$) we compare with the result obtained from the exact diagonalization of $H_{\rm eff}$, shown in stars. We observe that the Uimin-Lai-Sutherland term enhances the degeneracy within a triplet. Right: Splitting in the middle of the single-domain-wall spectrum, for the $N$th triplet [in a system of $N$ sites, $H_\text{1dw}$ is a $6N\times 6N$ matrix. The $N$th triplet contains the energy levels $E_{3N-2}=\lambda_{1}(N), E_{3N-1}=\lambda_2(N)$ and $E_{3N}=\lambda_{3}(N)$]. For $\Phi$ sufficiently away from $\pi/3$, the splitting decays exponentially with $N$.}
\label{fig:splitting}
\end{figure}

We next numerically diagonalize $H_\text{1dw}$ and study the dependence of the $\mathbb{Z}_3$-triplet splittings on the system size. 
We define the splitting as 
\begin{equation}
 {\rm Splitting}_n=\sqrt{(\lambda_1(n)-\lambda_{2}(n))^2+(\lambda_{2}(n)-\lambda_{3}(n))^2+(\lambda_{3}(n)-\lambda_{1}(n))^2},
\end{equation}
where $\lambda_{1,2,3}(n)$ are the energy eigenvalues of the $\mathbb{Z}_3$ triplet $n$.
The $\mathbb{Z}_3$-triplet splittings in the low-lying part and in the middle of the single-domain-wall spectrum are shown in Fig.~\ref{fig:splitting} left and right, respectively. 

We first focus on the low-lying sector in the nonchiral ferromagnetic limit $\Phi=\pi/3$, taking $J_\perp=0.1V'$ for concreteness. We observe that for $J=0$, the splittings decay subexponentially in $N$, consistently with the absence of strong zero modes in the nonchiral quantum clock model~\cite{Fendley2012,Jermyn2014,Alicea2016}
Remarkably, upon the inclusion of nonzero $H_{\rm S}$ with $J^2/2U>0.01V^\prime$, the decay becomes exponential, suggesting that the Uimin-Lai-Sutherland perturbation helps the development of the zero mode. This observation can be understood by noting that  $H_{\rm S}$, similarly to chirality, promotes the splitting between the bands of the domain-wall conjugacy classes. 
Namely, for periodic boundary conditions, the domain walls
$F_1=\{|{\bf A}|{\bf B}\rangle,|{\bf B}|{\bf C}\rangle,|{\bf C}|{\bf A}\rangle\}$ have band energies $E_+(k)=\frac{J^2}{2U}+V'\cos\Phi+2J_\perp\cos(k)$, while the other set of domain walls $F_2=\{|{\bf A}|{\bf C}\rangle,|{\bf B}|{\bf A}\rangle,|{\bf C}|{\bf B}\rangle\}$ has energies $E_-(k)=V'\cos(\Phi-\frac{2\pi}{3})+2J_\perp\cos(k)$. 
The processes that mix states within each family (taking the family $F_1$ for definitness) can occur only via:
\begin{itemize}
 \item A member of $F_1$ switching to a member of $F_2$ on one end of the chain, then
 \item the state in $F_2$ tunneling through the whole system to reach the opposite end, and finally
 \item the state in the end of the chain switching back to another member of $F_1$.
\end{itemize}
When the energy difference between the bottom of the higher-energy band and the top of the lower-energy band is greater than zero, i.e., 
\begin{equation}
 \Delta=E_+(\pi)-E_-(0)=\frac{J^2}{2U}-\sqrt{3}V^\prime\sin\left(\Phi-\frac{\pi}{3}\right)-4J_\perp>0
\end{equation}
for $\Phi\lesssim \pi/3$, this leads to an exponentially small (in the system size) matrix element between the states in each family~\cite{Fendley2012,Jermyn2014,Alicea2016}. Note that for $J^2/2U$ sufficiently larger than $J_\perp$, this holds even in the nonchiral $\Phi=\pi/3$ case. 

Finally, we study the splittings in the middle of the spectrum of $H_\text{1dw}$. 
While in this case we find that $H_{\rm S}$ alone does not introduce exponentially decaying splittings in the ferromagnetic limit, we also find that for fixed $J_\perp$ and $J$, moving with $\Phi$ away from the nonchiral $\Phi=\pi/3$ limit leads to a transition from a power-law to an exponential decay in the $\mathbb{Z}_3$-triplet splittings, suggesting that chiral interactions lead to the emergence of strong zero modes.

\begin{figure}[b!]
\includegraphics[scale=0.42]{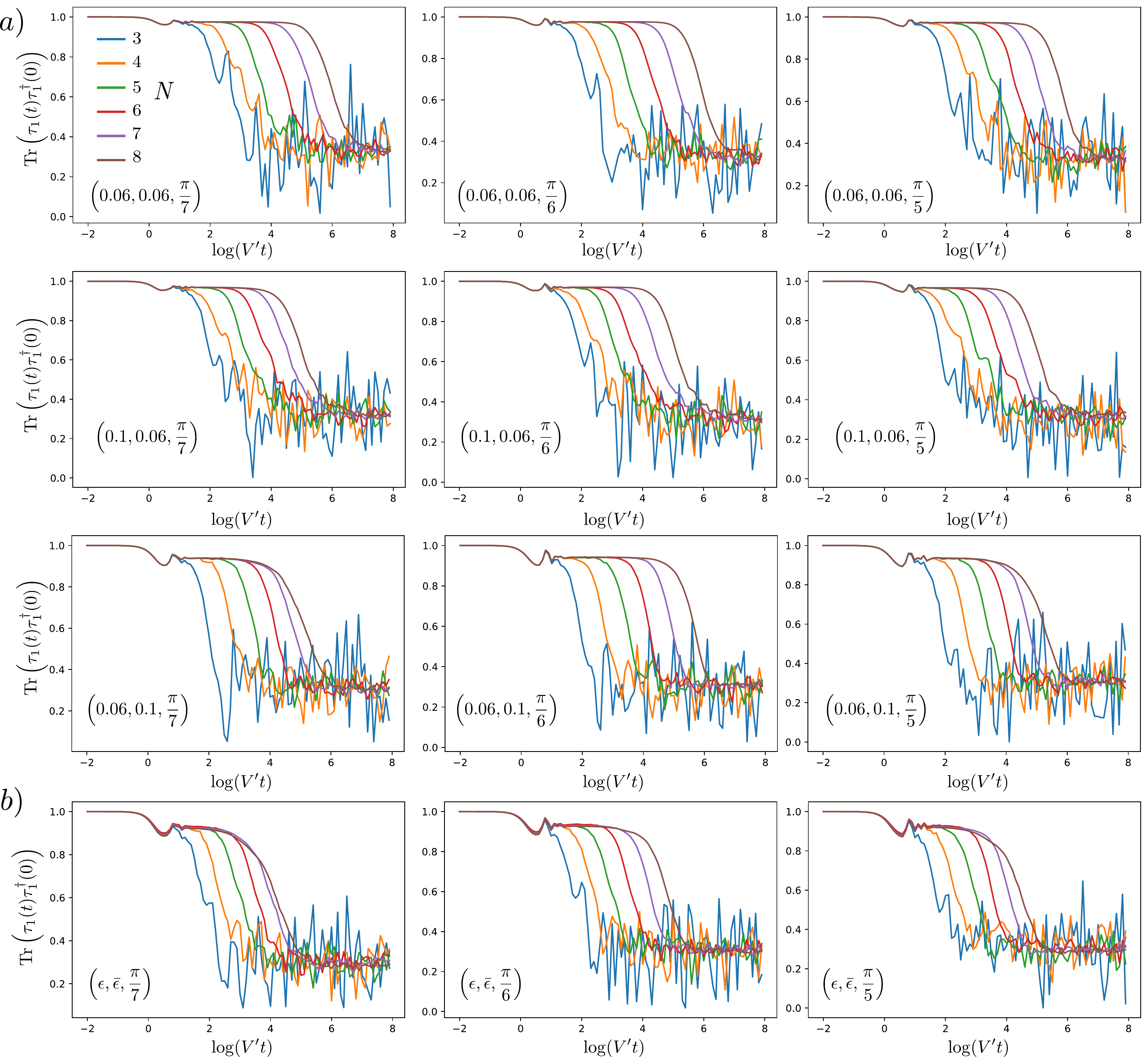}
    \caption{${\rm Tr}(\tau_1(t)\tau_1^\dagger(0))$ for different values of $\left(\frac{J^2}{2UV'},J_\perp/V',\Phi\right)$ and system sizes $N$, as function of dimensionless time $V^\prime t$. a.- Results for clean systems. b.- Results for disordered systems. 
    In the latter case, 
    we disorder the parameters using the random variables  $\epsilon$ and $\bar{\epsilon}$, which for each site are uniformly distributed on the interval $[0.1, 0.12]$ with no spatial correlations. 
}
 \label{fig:correlators}
\end{figure}

\newpage
\vspace*{-4em}

\subsection{Dynamical correlator}
\label{sec:dynamics}
\vspace*{-1em}

We complement the above results by examining the dynamical correlator ${\rm Tr}(\tau_1(t)\tau_1^\dagger(0))$ (defined in the main text) for various values of the parameters. 
This gives us a probe sensitive to the full spectrum of the system, not only the low-lying states. 
These correlators, computed using exact diagonalization of Eq.~\eqref{eq:Heff_5}, are shown in Fig.~\ref{fig:correlators}.

For a clean system (Fig.~\ref{fig:correlators}a), and weak $J^2/2U$, $J_\perp$,  we observe behavior consistent with our findings about the spectrum:
The parameter $J_\perp/V'$ is seen to be the main factor that influences by how much increasing $N$ increases the time for which the correlator persist near its initial value; the influence of $J^2/(2UV')$ appears to be less important (albeit not negligible), especially for $\Phi\gtrsim \pi/6$.  (This is the middle and the ``ferromagnetic" side of the chiral regime in the ``magnetic" nomenclature of Sec.~\ref{sec:1dw}.)

In the presence of disordered couplings (Fig.~ \ref{fig:correlators}b), our observations are similar: the results are consistent with the correlator remaining close to its initial value for times that increase exponentially with the system size, and hence with strong zero modes, especially for $\Phi\gtrsim \pi/6$. (Note that our disorder simulations are for slightly stronger $J^2/2U$ and $J_\perp$ couplings than those in the clean systems.)

\enlargethispage{2cm}

The results of Secs.~\ref{sec:spectral-signatures} and \ref{sec:dynamics} provide strong corroborating evidence that the appearance of the strong zero modes requires no fine tuning; the strong zero modes are present in a whole region of the parameter space.


\begin{thebibliography}{53}
\bibitem [{sym()}]{symmfn}%
  \BibitemOpen
  \href@noop {} {}\bibinfo {note} {The model is also invariant under reflection
  along the legs combined with exchanging two of the legs; this symmetry can be
  used instead of time-reversal in our considerations.}\BibitemShut {Stop}%
\bibitem [{kFf()}]{kFfn}%
  \BibitemOpen
  \href@noop {} {}\bibinfo {note} {The model's single-particle Hamiltonian
  $H_0$ has a reflection symmetry which leads to $k_{F,1}=k_{F,2}$; however,
  this is not a generic feature. All the processes we consider survive in the
  more general $\mathbb{Z}_3$- and time-reversal-invariant case with split
  $k_{F,1}$ and $k_{F,2}$.}\BibitemShut {Stop}%
\end{thebibliography}
\end{document}